\documentclass[aps,preprint,nofootinbib,superscriptaddress]{revtex4-1}

\usepackage[utf8]{inputenc}
\usepackage[T1]{fontenc}
\usepackage{amsmath,amssymb}
\usepackage{graphicx}
\usepackage{xcolor}
\usepackage{hyperref}
\usepackage{slashed}
\usepackage{cleveref}
\usepackage{braket}
\usepackage{tikz}
\usepackage{pgfplots}
\pgfplotsset{compat=1.17}

\newcommand{\ee}{\end{equation}}
\newcommand{\bb}{\begin{equation}}
\newcommand{\eqb}{\begin{eqnarray}}
\newcommand{\eqf}{\end{eqnarray}}

\begin{document}

\title{Non–Abelian Recoil Geometry and Infrared Holonomies in Heavy–Quark Transitions}

\author{J.~Gamboa}
\affiliation{Departamento de Física, Universidad de Santiago de Chile, Santiago, Chile}
\email{jorge.gamboa@usach.cl}

\author{N.~Tapia-Arellano}
\affiliation{Department of Physics and Astronomy, Agnes Scott College, Decatur, GA. 30030, USA}
\email{narellano@agnesscott.edu}

\begin{abstract}
We propose a geometric formulation of heavy-quark transitions in which infrared-dressed states are adiabatically transported in the multidimensional recoil space and acquire Berry holonomies.  Within this framework, single-step decays are governed by an abelian geometric phase and reproduce the standard Isgur-Wise behaviour, while sequential decays probe genuinely non-Abelian
holonomies associated with a two-dimensional recoil space.  The resulting geometric structure correlates different decay channels and provides a unified interpretation of mixing effects and quasi-degenerate states in heavy-quark phenomenology.  This approach suggests that several long-standing puzzles arise as geometric
consequences of infrared dressing rather than as accidental features of the microscopic dynamics.
\end{abstract}

\maketitle

\section{Introduction}
Sequential heavy-quark decays probe a qualitatively new regime of heavy-quark dynamics. Unlike single-step transitions, which depend on a single recoil variable, sequential processes involve two independent changes of the heavy-quark velocity
\cite{Isgur:1989vq,Neubert:1993mb,Georgi:1990um,Chay:1990da,Manohar:2000dt,
Mannel:1990vg,Ali:1991fy,Mannel:1991mc}.
As a result, the recoil space becomes intrinsically two-dimensional. In this work, we show that this simple kinematic fact enforces a non-Abelian Berry curvature in the infrared cloud of the heavy-light system, giving rise to an emergent $\mathrm{SU}(2)$ holonomy.

More generally, this geometric structure is not unique to heavy-quark sequential decays. It is a generic feature of physical systems that exhibit quasi-degenerate states and evolve in an adiabatic regime, in which slowly varying parameters induce non-Abelian geometric phases in the corresponding low-energy subspace.

A paradigmatic realization of this mechanism is provided by a sequential decay $A \to B \to C$, which involves two successive changes of the heavy-quark velocity characterized by the recoil parameters
\begin{equation}
w_1 = v_A\!\cdot v_B,
\qquad
w_2 = v_B\!\cdot v_C.
\end{equation}
These parameters are generically linearly independent, so that the recoil manifold is genuinely two-dimensional. The relevant quasi-degeneracy does not concern the energies of the external states $A$, $B$, and $C$, but rather arises within the internal heavy--light sector. For fixed $(w_1,w_2)$, two or more internal configurations with identical quantum numbers can become nearly
degenerate, defining a low-dimensional subspace in which adiabatic mixing takes place.

For such a doubly degenerate system, it is natural to introduce an effective Hamiltonian acting on the quasi-degenerate subspace \cite{Berry:1984jv},
\begin{equation}
H_{\rm eff}
=
E_0\,\mathbb{I}
+
\boldsymbol{\sigma}\!\cdot\!\mathbf{B},
\end{equation}
which is formally analogous to the Hamiltonian of a spin--$\tfrac12$ particle in an effective magnetic field. Here, the vector $\mathbf{B}$ does not represent a physical magnetic field, but an adiabatic field generated by the slow variation of the external parameters. In the present case, these parameters are precisely the recoil variables $(w_1,w_2)$, which evolve slowly in the near--zero--recoil regime $w_i-1\ll1$.

The explicit form of $\mathbf{B}$ follows from the adiabatic approximation and is encoded in the Berry curvature, which plays the role of a non-Abelian magnetic field in parameter space. It is given by \cite{Wilczek:1984dh}
\begin{equation}
\mathcal{F}
=
\nabla \times \mathcal{A}
+
\mathcal{A}\times\mathcal{A},
\end{equation}
where $\mathcal{A}$ is the Berry connection taking values in the Lie algebra $\mathfrak{u}(2)$ associated with the quasi--degenerate subspace. This non-Abelian curvature is the ultimate origin of the $\mathrm{SU}(2)$ holonomy that governs the adiabatic mixing of the near-degenerate states.

The same geometric mechanism also underlies the dynamics of near-threshold exotic hadronic states \cite{Brambilla:2019esw,Aaij:2013zoa}. In such systems, the role of the recoil variables is played by internal channel–mixing parameters, such as isospin,
which control the adiabatic evolution within a low-dimensional space of nearly degenerate hadronic configurations. Large isospin admixtures and the
emergence of exotic states, therefore, arise as manifestations of the same non-Abelian infrared geometry revealed in sequential heavy-quark decays.

The aim of this paper is to analyze, in a general setting, the dynamics of quasi-degenerate systems evolving in an adiabatic regime and the associated
emergence of non-Abelian geometric structures. We then apply this framework to sequential heavy-quark decays, proposing a geometric resolution of the
long-standing $j=\tfrac12$ versus $j=\tfrac32$ puzzle and providing an interpretation of large isospin mixing in terms of infrared holonomies. In the
conclusions, we discuss the broader physical implications of our results and their natural application to near-threshold exotic states such as the
$X(3872)$.

\section{Sequential Decays and Non-Abelian Recoil Geometry}

We begin with the simplest heavy-to-heavy transition, namely the single-step decay
\[
   B \to D^{*},
\]
which depends on a single recoil variable and therefore probes an effectively abelian infrared geometry. This process plays a distinguished role in
Heavy-quark physics as the prototypical heavy-to-heavy transition governed by a single recoil parameter
\cite{Isgur:1989vq,Neubert:1993mb,Georgi:1990um,Chay:1990da,Manohar:2000dt,
Mannel:1990vg,Ali:1991fy,Mannel:1991mc}.
In the standard Heavy Quarks Effective Theory (HQET), this decay is computed under the assumption that the physical asymptotic states belong to the Fock space, so that all infrared effects are encoded in a single universal Isgur-Wise function.

This assumption reflects the adiabatic nature of the heavy-quark limit: non-adiabatic Born-Oppenheimer corrections are suppressed by $1/m_Q$, and the
heavy-light system is therefore usually treated as evolving without \emph{explicit} geometric phases, with physical states identified as Fock-space asymptotic states. If such corrections are not entirely negligible, however, a more appropriate
description involves \emph{infrared--dressed} states in the sense of Chung--Kibble--Kulish--Faddeev (CKKF)
\cite{Chung:1965zza,Kibble:1968sfb,Kulish:1970ut}, supplemented by a Berry phase of geometric origin that encodes the adiabatic response of the infrared dressing
to changes in the heavy-quark velocity.

In this framework, the relevant slow parameter is the relative angle $\gamma$ between the initial and final heavy-quark velocities, defined through
\begin{equation}
v\!\cdot\!v' = \cosh\gamma ,
\end{equation}
which provides a natural geometric coordinate in recoil space.

This single-recoil process will serve as a reference point for the genuinely new geometric features that arise in sequential decays. As we shall see, when
the heavy--quark transition proceeds through successive recoil steps, the recoil space becomes multi-dimensional, and the associated infrared geometry
is no longer abelian, giving rise to a non-Abelian Berry curvature.

Within this adiabatic and infrared-dressed framework
\cite{Gamboa:2025dry,Gamboa:2025fcn,Gamboa:2025nco,Gamboa:2025qjr}, the transition amplitude acquires a natural geometric interpretation. The initial
and final heavy-light states are described by two dressing holonomies—path-ordered exponentials of the Berry connection—associated with the adiabatic trajectories traced by the heavy-quark velocities $v$ and $v'$. The corresponding hadronic matrix element can then be written as
\[
   \Xi_{\rm geom}(w)
   =
   \big\langle 0 \big|\,
      \mathcal{U}_{C}^{\dagger}(v')\,\mathcal{U}_{C}(v)
   \big| 0 \big\rangle ,
\]
which reduces smoothly to the Isgur-Wise function in the heavy-quark limit. Its normalization at zero recoil and its slope are fixed by the geometry of the
Berry connection. For the process $B\to D^{*}$, the overlap of the two holonomies is well approximated by an exponential form, naturally reproducing the observed universality of the differential rate $d\Gamma/dw$.

\subsection{Non--Abelian Recoil Processes}

A qualitatively distinct regime is encountered in the sequential decay
\cite{Isgur:1989vq,Isgur:1990jf}
\begin{equation}
   B \to D^{**} \to D^{*}.
\end{equation}
In contrast to direct heavy-quark transitions, which depend on a single recoil parameter, this process involves two successive changes of the heavy-quark
velocity. The corresponding recoil variables $(w_1,w_2)$ therefore parametrize a genuinely two-dimensional region in velocity space. On this surface, the Berry curvature no longer vanishes, leading to an emergent $\mathrm{SU}(2)$ geometric structure that has no counterpart in conventional HQET.

The kinematics of the process define a piecewise adiabatic trajectory in velocity space,
\[
   v \;\longrightarrow\; v' \;\longrightarrow\; v'',
\]
associated with two consecutive recoil steps. As a result, the infrared dressing factors accumulated along each segment do not combine abelianly.
Instead, the corresponding holonomies obey the non-Abelian composition rule
\[
   \mathcal{U}(v''\!\leftarrow v')\,\mathcal{U}(v'\!\leftarrow v)
   \neq
   \mathcal{U}(v''\!\leftarrow v),
\]
which directly reflects the presence of a nonvanishing Berry curvature on the two-dimensional recoil manifold.

Close to the zero--recoil point, the response of the infrared cloud to each independent velocity variation can be characterized by two slope operators,
\[
   R_{1} = \vec r_{1}\!\cdot\!\vec\sigma,
   \qquad
   R_{2} = \vec r_{2}\!\cdot\!\vec\sigma,
\]
associated with the two recoil directions. Their algebra encodes the local curvature of the recoil surface,
\[
   [R_{1},R_{2}] = 2i\,(\vec r_{1}\times\vec r_{2})\!\cdot\!\vec\sigma,
\]
which vanishes only when the two velocity changes are aligned.

In this regime, the non-Abelian holonomy generated by the sequential recoil takes the approximate form
\[
   \hat{\Xi}(w_{1},w_{2})
   \simeq
   e^{-(w_{1}-1)R_{1} - (w_{2}-1)R_{2}},
\]
with eigenvalues
\[
   \Xi_{\pm}(w_{1},w_{2})
   =
   \exp\!\big[\mp\,|\vec\alpha(w_{1},w_{2})|\big],
   \qquad
   \vec\alpha :=
   (w_{1}-1)\vec r_{1} + (w_{2}-1)\vec r_{2}.
\]

For later convenience, we introduce the parametrization
\[
\rho_1 = |\vec r_{1}|,\qquad
\rho_2 = |\vec r_{2}|,\qquad
\cos\chi = \frac{\vec r_{1}\!\cdot\!\vec r_{2}}{\rho_1\rho_2},
\]
which yields
\[
|\vec\alpha|^{2}
=
\rho_1^{2}(w_1-1)^{2}
+
\rho_2^{2}(w_2-1)^{2}
+
2\rho_1\rho_2\cos\chi\,(w_1-1)(w_2-1).
\]
The two eigenvalues of the holonomy, therefore, define a pair of universal geometric Isgur--Wise functions,
\[
\Xi_{\pm}(w_1,w_2)
=
\exp\!\left[
   \mp\sqrt{
      \rho_1^{2}(w_1-1)^{2}
      + \rho_2^{2}(w_2-1)^{2}
      + 2\rho_1\rho_2\cos\chi\,(w_1-1)(w_2-1)
   }
\right].
\]

As will become clear in the remainder of this work, all physical form factors associated with sequential heavy-quark decays can be obtained as channel-dependent projections of these two universal modes. The functions $\Xi_{\pm}(w_1,w_2)$ should therefore be viewed not as hadronic form factors themselves, but as the eigenmodes of the underlying $\mathrm{SU}(2)$ Berry
holonomy. Observable Isgur-Wise functions arise from projecting these modes onto specific heavy-light states, with weights fixed by their orientation in the
internal $\mathrm{SU}(2)$ space.

\subsection{Quasi--Degeneracy, Adiabatic Mixing, and Exotic States}

An apparently distinct but deeply related realization of the same geometric mechanism arises in the physics of exotic hadronic states. In this case, the role of the recoil variables $(w_1,w_2)$ is not played by changes of the heavy-quark velocity, but by slow parameters controlling the mixing of hadronic channels with identical quantum numbers, such as isospin or
near--threshold configurations.

When two such channels lie close in energy and are dominated by near-threshold dynamics, the effective state space becomes two-dimensional, defining a quasi-degenerate subspace analogous to that encountered in sequential heavy-quark decays. The evolution of the system through this region is adiabatic, and the physical state cannot be identified with a single microscopic configuration. Instead, it emerges as an adiabatic eigenstate selected within this low-dimensional subspace.

From a technical point of view, the mathematical structure is identical to the one discussed in the previous subsection. The same non-Abelian Berry connection, curvature, and $\mathrm{SU}(2)$ holonomy govern the adiabatic transport within the quasi-degenerate subspace. Different physical channels correspond to different projections of the same two universal geometric eigenmodes $\Xi_{\pm}$.

As in the sequential decay $B\to D^{**}\to D^{*}$, the observable amplitudes are therefore not independent dynamical objects, but correlated projections of a single underlying infrared holonomy. In this sense, the structure uncovered in sequential heavy-quark decays provides a unifying geometric framework for understanding the formation and mixing of near-threshold exotic states.

% ============================
%  Puzzle
% ============================
\noindent
\section{The $\tfrac{3}{2}$ vs.\ $\tfrac{1}{2}$ Puzzle}

A key application of the geometric framework is a geometric resolution of the long-standing $3/2$ versus $1/2$ puzzle \cite{Leibovich:1997tu,Leibovich:1997em}.
Orbitally excited charmed mesons ($L=1$) form two heavy-quark doublets with $j_\ell=\tfrac12$ and $j_\ell=\tfrac32$. In standard HQET, where the two channels are treated as independent dynamical amplitudes, one expects
\[
   \Gamma_{3/2} \gg \Gamma_{1/2},
\]
whereas experimentally, one finds instead
\cite{CLEO:1997thi,BaBar:2008cfy,BaBar:2008dar,Belle:2007uwr,ALEPH:1996qht,
Dracos:2000tu,OPAL:2002yqc,LHCb:2019juy}
\[
   \Gamma_{3/2}\sim\Gamma_{1/2}.
\]
In HQET the inequality $\Gamma_{3/2}\gg\Gamma_{1/2}$ arises because the two $L=1$ doublets are described by \emph{a priori} independent hadronic amplitudes (or, equivalently, independent Isgur--Wise functions), whose normalizations and slopes are only weakly constrained beyond sum rules \cite{Bjorken:1991kg,Uraltsev:2000ce}.
In the geometric framework, this independence is lost: both doublets probe the same infrared holonomy, so their form factors become correlated projections of the same universal geometric eigenmodes. The observed pattern, therefore, reflects a geometric constraint rather than channel-dependent tuning.

In the geometric approach developed above, the relevant hadronic dynamics are encoded in the non-Abelian holonomy $\hat{\Xi}$ associated with the infrared cloud. Let $\{|+\rangle,|-\rangle\}$ denote its eigenbasis.
The physical $L=1$ doublets do not coincide with this geometric basis, but are related to it by a single rotation angle $\gamma$,
\[
   |j_\ell=\tfrac12\rangle
   = \cos\frac{\gamma}{2}\,|+\rangle
     + \sin\frac{\gamma}{2}\,|-\rangle,
\qquad
   |j_\ell=\tfrac32\rangle
   = -\sin\frac{\gamma}{2}\,|+\rangle
     + \cos\frac{\gamma}{2}\,|-\rangle.
\]
As a consequence, the physical form factors are not independent functions, but fixed linear combinations of the same two universal geometric eigenmodes,
\[
   F_{1/2}
   = \cos^{2}\!\frac{\gamma}{2}\,\Xi_{+}
     + \sin^{2}\!\frac{\gamma}{2}\,\Xi_{-},
\qquad
   F_{3/2}
   = \sin^{2}\!\frac{\gamma}{2}\,\Xi_{+}
     + \cos^{2}\!\frac{\gamma}{2}\,\Xi_{-}.
\]

The partial widths follow from integrating the squared form factors over the
allowed kinematic manifold. We write them compactly as
\[
   \Gamma_{1/2} \;\propto\;
      A_{1/2,+}\,I_{+} + A_{1/2,-}\,I_{-},
\qquad
   \Gamma_{3/2} \;\propto\;
      A_{3/2,+}\,I_{+} + A_{3/2,-}\,I_{-},
\]
with
\[
   I_{\pm}
   =
   \int d\Phi(w_{1},w_{2})\,|\Xi_{\pm}(w_{1},w_{2})|^{2}.
\]
The phase--space measure $d\Phi(w_{1},w_{2})$ is purely kinematic and contains no hadronic input; all dynamical information resides in the universal geometric modes $\Xi_{\pm}$.

The crucial point is that the two channels share the same geometric integrals $I_\pm$: the only difference between $j_\ell=\tfrac12$ and $j_\ell=\tfrac32$ is the complementary weighting of $I_+$ and $I_-$ fixed by the single mixing angle $\gamma$. A large parametric hierarchy can therefore arise only if either $I_+\!\gg\! I_-$ or $I_-\!\gg\! I_+$, which is not expected in the presence of a non-Abelian $\mathrm{SU}(2)$ holonomy. Instead, for $I_+\sim I_-$, one naturally finds
\[
   \Gamma_{1/2}\sim\Gamma_{3/2},
\]
in agreement with experiment.

The key point is thus geometric: in any sequential heavy-quark process, the recoil space is intrinsically two-dimensional. This enforces a non-commutative
Berry phase and an emergent $\mathrm{SU}(2)$ holonomy \cite{Wilczek:1984dh}, so that the $j_\ell=\tfrac12$ and $j_\ell=\tfrac32$ channels are not independent dynamical objects, but different projections of the same two universal geometric eigenmodes. The empirical similarity of their partial widths, therefore, reflects the infrared geometric organization of the brown-muck response rather than a breakdown of heavy-quark symmetry.

Importantly, heavy-quark symmetry itself is not modified: the new ingredient is the infrared geometric organization of the brown-muck response, which correlates channels that HQET treats as independent at leading order.

\section{On the Large Isospin Mixing}

A closely related application of the geometric framework concerns the problem of large isospin mixing in near-threshold hadronic systems. In conventional
hadronic descriptions, channels with different isospin quantum numbers are often treated as independent, up to small symmetry-breaking effects induced by quark mass differences or electromagnetic interactions. As a consequence, large isospin mixing is usually regarded as accidental or fine-tuned.

Within the geometric framework developed in this work, this viewpoint is naturally revised. When two hadronic channels with identical conserved quantum numbers but different isospin, such as $I=0$ and $I=1$, lie close in energy and are probed in an adiabatic near-threshold regime, they form a quasi-degenerate two-dimensional subspace. The relevant dynamics are then
governed by the same non-Abelian infrared holonomy that controls sequential heavy-quark decays.

Let $\{|I=0\rangle, |I=1\rangle\}$ denote a basis of isospin channels in this quasi-degenerate subspace. As in the $j_\ell=\tfrac12$ versus $j_\ell=\tfrac32$ case, these physical channels do not coincide with the
eigenbasis $\{|+\rangle,|-\rangle\}$ of the geometric holonomy $\hat{\Xi}$.
The two bases are instead related by a single mixing angle $\theta$, which parametrizes the relative orientation between the physical isospin basis and
the adiabatic eigenbasis selected by the infrared holonomy,
\[
   |I=0\rangle
   = \cos\frac{\theta}{2}\,|+\rangle
     + \sin\frac{\theta}{2}\,|-\rangle,
\qquad
   |I=1\rangle
   = -\sin\frac{\theta}{2}\,|+\rangle
     + \cos\frac{\theta}{2}\,|-\rangle.
\]

As a result, the corresponding physical amplitudes are again correlated projections of the same universal geometric eigenmodes,
\[
   F_{I=0}
   = \cos^{2}\!\frac{\theta}{2}\,\Xi_{+}
     + \sin^{2}\!\frac{\theta}{2}\,\Xi_{-},
\qquad
   F_{I=1}
   = \sin^{2}\!\frac{\theta}{2}\,\Xi_{+}
     + \cos^{2}\!\frac{\theta}{2}\,\Xi_{-}.
\]

Observable decay rates or production probabilities follow from integrating the squared amplitudes over the appropriate phase space. As in the heavy-quark case, the two isospin channels share the same geometric integrals $I_\pm$,
\[
   I_{\pm}
   =
   \int d\Phi\,|\Xi_{\pm}|^{2},
\]
and differ only by the complementary weighting determined by the mixing angle $\theta$. Large isospin mixing, therefore, arises naturally whenever the geometric integrals are comparable, $I_+\sim I_-$, and does not require any fine-tuning of microscopic dynamics.

From this perspective, large isospin violation is not a symptom of symmetry breaking at short distances, but rather a manifestation of the infrared geometric organization of the near-threshold state space.

The geometric interpretation developed in this work provides a unified framework for understanding a variety of seemingly unrelated puzzles in hadronic physics. Both the $j_\ell=\tfrac12$ versus $j_\ell=\tfrac32$ puzzle in sequential heavy-quark decays and the appearance of large isospin mixing in
near-threshold systems are shown to originate from the same underlying mechanism: adiabatic evolution within a quasi-degenerate two-dimensional subspace, governed by a non-Abelian $\mathrm{SU}(2)$ holonomy of the infrared cloud.

A paradigmatic example is provided by the $X(3872)$, which lies extremely close to the $D^0\bar D^{*0}$ threshold and exhibits strong isospin mixing. In the
present framework, this behavior is not accidental. Rather, the $X(3872)$ emerges as an adiabatic eigenstate selected by the infrared geometry, with its
observed isospin composition reflecting a particular projection of the same universal geometric eigenmodes that control sequential heavy-quark transitions.

From this viewpoint, exotic states such as the $X(3872)$ are not characterized by a specific microscopic structure, but by their position within the infrared geometric landscape of QCD. The framework developed here thus suggests that near-threshold exotics, heavy-quark decay puzzles, and large isospin mixing are unified manifestations of a common non-Abelian infrared organization.

\section{Conclusions and Outlook}

In this letter, we have shown that heavy-quark decays admit a natural geometric description once infrared dressing and Berry phases are taken into account.
Single-step transitions such as $B\!\to\!D^{*}$ are governed by an abelian holonomy and reproduce the standard Isgur-Wise behaviour, while sequential
decays probe a genuinely two-dimensional recoil space and give rise to a non-Abelian $\mathrm{SU}(2)$ Berry holonomy \cite{Wilczek:1984dh}.

Within this framework, the $j_\ell=\tfrac12$ and $j_\ell=\tfrac32$ channels are not independent dynamical amplitudes, but different projections of the same two universal geometric eigenmodes. As a result, the partial widths of the two $L=1$ doublets are naturally correlated, providing a geometric explanation of the observed $1/2$ versus $3/2$ pattern without violating heavy--quark symmetry or established sum rules.

The same geometric viewpoint also leads to a natural reinterpretation of large isospin mixing in near-threshold hadronic systems. When channels with different isospin quantum numbers become quasi-degenerate and are probed in an adiabatic regime, their mixing is governed by the same non-Abelian infrared geometry. Large isospin admixtures therefore, emerge as geometric projections of universal infrared eigenmodes, rather than as accidental consequences of
short--distance symmetry breaking.

From this perspective, the $1/m_Q$ corrections of HQET acquire a clear physical interpretation as controlled departures from adiabatic transport in the space of infrared configurations, analogous to non-adiabatic corrections in Born-Oppenheimer systems. A systematic analysis of this non-adiabatic sector lies beyond the scope of the present Letter, but the results presented here suggest that infrared geometry provides a powerful and unifying organizing principle for heavy-quark dynamics and for the structure of near-threshold
exotic states.

\acknowledgments
\noindent
 This research was supported by DICYT (USACH), grant number 042531GR\_REG. The work of N.T.A is supported by Agnes Scott College.

%===========================
% REFERENCES
%===========================

\bibliographystyle{JHEP}
\bibliography{ref.bib}

\end{document}